\documentclass[aps,prl,twocolumn,superscriptaddress,longbibliography,nofootinbib,secnumarabic]{revtex4-2}
\usepackage[T1]{fontenc}
\usepackage[utf8]{inputenc}
\usepackage[english]{babel}   
\usepackage{lmodern}
\usepackage{amsmath,amssymb,amsthm,bm,bbm}
\usepackage{siunitx}
\usepackage{xspace}
\usepackage{graphicx}
\usepackage{hyperref}
\usepackage{braket}
\usepackage{physics}        
\usepackage{enumitem}       
\usepackage{times}
\usepackage{siunitx}  
\usepackage{listings}
\usepackage{xcolor} 
\usepackage{mathtools }
\usepackage{tabularx}
\usepackage{tabularray}
\usepackage{threeparttable}
\usepackage{orcidlink} 
\usepackage{tcolorbox}
\usepackage{booktabs}

\hypersetup{
  colorlinks=true,
  linkcolor=magenta,
  citecolor=magenta,
  urlcolor=magenta
}

\newtcolorbox{draftrefs}{
  colback=gray!10,
  colframe=gray!50,
  title={References (for this section --- to verify later)},
  fonttitle=\bfseries
}

\begin{document}

\title{Self-Generated Chiral Rotation in Whispering-Gallery Optomechanics}

\author{Mohamed Hatifi\,\orcidlink{0009-0005-3368-2751}}
\affiliation{Aix Marseille Univ, CNRS, Centrale M\'editerran\'ee, Institut Fresnel, Marseille, France}

\email{hatifi@fresnel.fr}

\begin{abstract}
Backscattering in whispering-gallery-mode resonators is usually a passive mode-splitting mechanism produced by a fixed defect. Here, we show that, when the backscatterer is a mechanical angular degree of freedom, the same process becomes an angular-recoil backaction channel capable of generating chirality under reciprocal driving. A localized movable scatterer coherently converts photons between clockwise and counterclockwise whispering-gallery modes, transferring angular recoil in each circulation-changing event. In a weak-scattering driven-dissipative model, reciprocal bidirectional pumping gives zero net torque at rest, but rotation Doppler-shifts the two opposite scattering rates in opposite directions. For suitable detuning, this feedback produces negative angular friction, destabilizes the nonrotating reciprocal state, and selects one of two symmetry-related steady rotations. The threshold scales inversely with the square of the WGM azimuthal index. The mechanically chiral state produces a direction-dependent weak-probe response, visible as a Doppler splitting of the backscattered spectra, turning passive WGM mode splitting into a minimal mechanism for autonomous chiral optomechanics.
\end{abstract}

\maketitle

\paragraph{Introduction.---}
Whispering-gallery-mode (WGM) resonators confine light in counterpropagating circulation states with large azimuthal angular momentum~\cite{allen1992,vahala2003,he2013,foreman2015}. When a localized object scatters the photon into the opposite circulation state, the optical angular momentum changes by twice that amount. In the usual fixed-defect setting, the corresponding recoil is absorbed by the support and appears optically as coherent mode splitting~\cite{gorodetsky2000,mazzei2007}. If the scatterer itself is a movable angular degree of freedom, the same circulation-changing event becomes a mechanical torque. The central question is whether reciprocal optical driving can make an initially achiral WGM system select a preferred direction of mechanical rotation without any externally imposed handedness. Existing work has established closely related mechanisms in different limits. Cavity optomechanics shows how drive and loss convert radiation pressure into dynamical backaction, amplification, and self-oscillation~\cite{kippenberg2005,aspelmeyer2014b,chesi2015,liu2026}. In WGM resonators, localized Rayleigh scatterers coherently couple clockwise and counterclockwise modes and generate standing-wave doublets~\cite{gorodetsky2000,mazzei2007}, while optical forces can transport nanoparticles along the resonator, as in WGM-carousel dynamics~\cite{arnold2009,vartabikashanian2026}. Direction-dependent WGM responses have also been obtained by externally spinning resonators, where Sagnac--Fizeau shifts make the counterpropagating modes inequivalent~\cite{maayani2018,jing2018}. In nonlinear ring resonators, bidirectional Kerr driving can spontaneously select an optical circulation imbalance~\cite{delbino2017,coillet2019}. These results provide optical backaction, WGM mode coupling, optical transport, imposed rotation, and optical symmetry breaking, but they leave open the recoil dynamics of a movable WGM backscatterer. The remaining case is a reciprocal WGM doublet in which the scatterer angle is a mechanical coordinate. In mode-splitting theories, the scatterer position is a fixed parameter. In one-sided transport, the drive already supplies a propagation direction. In spinning-resonator schemes, the angular velocity is externally prescribed. In Kerr symmetry-breaking, the selected order parameter is the optical circulation. Here, the possible order parameter is instead the mechanical velocity, and any handedness must emerge from angular-momentum exchange between backscattered photons and the scatterer itself. This geometric control of an optical phase is related in spirit to waveguide-QED settings where emitter positions tune reservoir phases and memory~\cite{sheremet2023,hatifi2026}; in the present WGM setting, the phase is attached to a recoiling mechanical coordinate. Here we show that this minimal setting realizes an angular-recoil form of optomechanical backaction. The scatterer position fixes the phase of the clockwise--counterclockwise scattering amplitude, so each circulation-changing event transfers a definite angular recoil to the mechanical coordinate. When the scatterer rotates, the two reciprocal scattering processes acquire opposite rotational Doppler shifts. For suitable detuning, this Doppler imbalance makes the optical torque act as negative angular friction, destabilizes the nonrotating reciprocal state, and produces two symmetry-related steady rotations, in analogy with autonomous mechanical-gain instabilities driven by dynamical backaction~\cite{kippenberg2005,aspelmeyer2014b,hatifi2025a,hatifi2026a}. We derive the recoil torque, the instability threshold, its inverse-square scaling with WGM index, and the direction-dependent weak-probe response of the rotating state. The result turns familiar WGM mode splitting from a passive signature of a static defect into a minimal mechanism for autonomous chiral optomechanics.

\paragraph{Angular-recoil coupling.---}
The phase of a localized WGM backscattering event fixes the minimal model. We consider a single optical doublet formed by two nearly degenerate counterpropagating modes, \(a_+\) and \(a_-\), with angular dependence \(e^{+im\theta}\) and \(e^{-im\theta}\). We refer to these as the clockwise (CW) and counterclockwise (CCW) modes. A scatterer at angular position \(\phi\) samples the local overlap of these fields; hence, the amplitude that converts a clockwise photon into a counterclockwise photon carries the phase \(e^{2im\phi}\), while the reverse process carries the conjugate phase \(e^{-2im\phi}\)~\cite{mazzei2007,deych2009}. If the scatterer is allowed to move, \(\phi\) becomes a mechanical coordinate conjugate to \(L_\phi\), and the conservative Hamiltonian reads
\begin{equation}
H =
\hbar\omega_c N
+
\frac{L_\phi^2}{2I}
+
V(\phi)
+
H_{\rm int},
\quad
N=a_+^\dagger a_+ + a_-^\dagger a_- ,
\label{eq:H_total}
\end{equation}
with \([\phi,L_\phi]=i\hbar\), moment of inertia \(I\), and
\begin{equation}
H_{\rm int}
=
\hbar J
\left(
e^{2im\phi}a_-^\dagger a_+
+
e^{-2im\phi}a_+^\dagger a_-
\right).
\label{eq:H_int}
\end{equation}
\begin{figure}[t]
\centering
\includegraphics[width=\columnwidth]{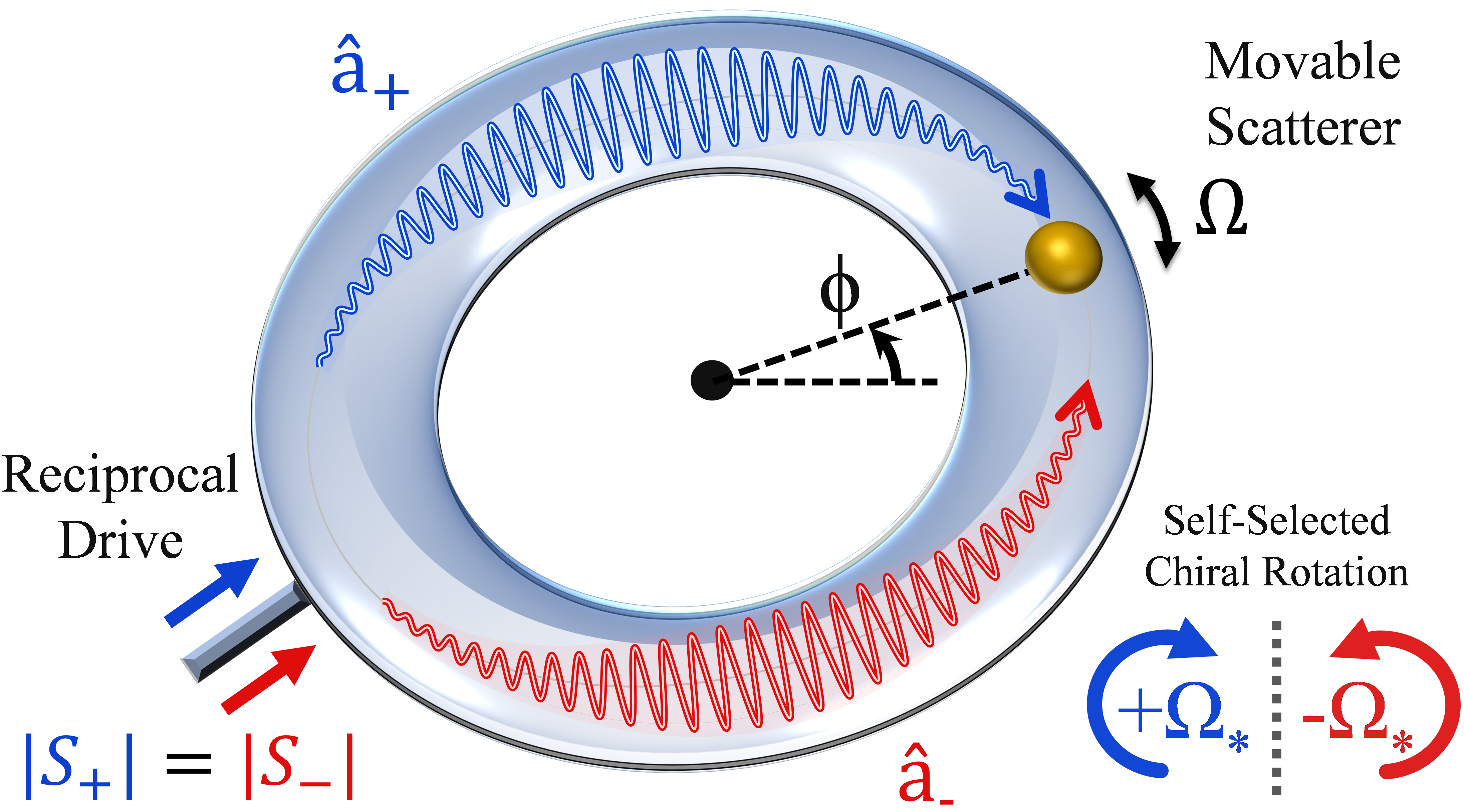}
\caption{
\textbf{Schematic of angular-recoil backaction in a whispering-gallery-mode resonator.} 
A localized movable scatterer at angular coordinate \(\phi\) couples the two counterpropagating WGM modes \(a_+\) and \(a_-\), which are driven reciprocally with equal inputs in amplitude \(|S_+|=|S_-|\). A circulation-changing scattering event transfers angular momentum to the scatterer: \(a_+\to a_-\) gives \(\Delta L_\phi=+2m\hbar\), while \(a_-\to a_+\) gives \(\Delta L_\phi=-2m\hbar\). If the scatterer rotates with angular velocity \(\Omega\), the two scattering channels acquire opposite rotational Doppler shifts, \(\Delta\mp2m\Omega\). This Doppler imbalance changes the two recoil rates, producing a net torque on the scatterer. Above threshold, the reciprocal state \(\Omega=0\) becomes unstable and the system selects one of two symmetry-related chiral rotations, \(\Omega=\pm\Omega_*\).
}
\label{fig:schematic}
\end{figure}
Here \(J\) is the coherent CW--CCW backscattering rate. Equations~\eqref{eq:H_total} and \eqref{eq:H_int} define the conservative angular-recoil coupling; optical driving and loss are introduced below at the Langevin/input-output level. For fixed \(\phi\), Eq.~\eqref{eq:H_int} is the usual scatterer-induced coupling responsible for WGM mode splitting~\cite{gorodetsky2000,mazzei2007,deych2009}. For dynamical \(\phi\), the same term becomes optomechanical; the mechanical coordinate controls the phase of a circulation-changing process rather than only shifting a resonance frequency~\cite{law1995,aspelmeyer2014b}. The recoil follows directly from the angular momentum carried by the two circulation states,
\begin{equation}
L_{\rm opt}
=
\hbar m
\left(
a_+^\dagger a_+
-
a_-^\dagger a_-
\right).
\label{eq:Lopt}
\end{equation}
The operator \(a_-^\dagger a_+\) changes the optical angular momentum by \(-2m\hbar\), while its Hermitian conjugate changes it by \(+2m\hbar\). The phase factors in Eq.~\eqref{eq:H_int} supply the opposite change to the mechanical angular momentum. Thus, each circulation-changing event transfers $\Delta L_\phi = \pm 2m\hbar$ to the scatterer~\cite{nieto-vesperinas2015}. Equivalently, the optical torque obtained from the Heisenberg equation for \(L_\phi\) is
\begin{equation}
\tau_{\rm opt}
=
\dot L_\phi
=
-2im\hbar J
\left(
e^{2im\phi}a_-^\dagger a_+
-
e^{-2im\phi}a_+^\dagger a_-
\right).
\label{eq:torque_operator}
\end{equation}
For the mean-field dynamics used below, \(a_\pm\rightarrow\alpha_\pm\), this becomes
\begin{equation}
\tau_{\rm opt}
=
4m\hbar J\,
\operatorname{Im}
\left[
e^{2im\phi}\alpha_-^*\alpha_+
\right].
\label{eq:classical_torque}
\end{equation}
The torque is therefore set by the phase-sensitive coherence between the two circulation states. Its magnitude contains the large WGM index \(m\), while its sign depends on the relative phase between the optical backscattering amplitude and the scatterer position. This instantaneous torque is the angular-recoil form of optomechanical backaction. Below, we evaluate its reciprocal-drive, time-averaged component for uniform rotation.

\paragraph{Reciprocal backaction.---}
The recoil torque becomes irreversible only because the scatterer is embedded in a driven, lossy optical resonator. We therefore supplement the conservative Hamiltonian by optical inputs and damping. In a frame rotating at the pump frequency \(\omega_L\), the optical Hamiltonian contains \(-\hbar\Delta N\), with $\Delta=\omega_L-\omega_c$ and $\gamma=\kappa/2$. Here \(\kappa\) represents the total optical linewidth. For coherent pump components, the rotating-frame drive term can be written
\begin{equation}
H_{\rm drv}
=
i\hbar
\sum_{\sigma=\pm}
\left(
S_\sigma a_\sigma^\dagger
-
S_\sigma^*a_\sigma
\right),
\label{eq:H_drive}
\end{equation}
where \(S_\sigma=\sqrt{\kappa_{\rm ex}}\,s_{{\rm in},\sigma}\) in the standard input-output normalization~\cite{gardiner1985,clerk2010b}. Including optical damping, the mean-field equations are
\begin{align}
\dot\alpha_+
&=
(i\Delta-\gamma)\alpha_+
-iJ e^{-2im\phi}\alpha_-
+S_+,
\label{eq:alpha_plus}
\\
\dot\alpha_-
&=
(i\Delta-\gamma)\alpha_-
-iJ e^{2im\phi}\alpha_+
+S_-,
\label{eq:alpha_minus}
\\
I\ddot\phi+\Gamma_\phi\dot\phi
&=
-\partial_\phi V
+
4m\hbar J\,
\operatorname{Im}
\left[
e^{2im\phi}\alpha_-^*\alpha_+
\right],
\label{eq:mechanical_eq}
\end{align}
where \(\Gamma_\phi\) is the angular damping coefficient. We use these equations in the weak-scattering, adiabatic regime
\begin{equation}
J\ll\gamma,
\qquad
\text{All mechanical rates}\ll\gamma,\nonumber
\label{eq:regime}
\end{equation}
so that the optical fields follow the scatterer's instantaneous angular velocity~\cite{aspelmeyer2014b}. This adiabatic condition concerns the time scale over which \(\Omega\) changes; the Doppler shift \(2m\Omega\) itself is retained explicitly and need not be small compared with \(\gamma\). Static pinning is first neglected, \(\partial_\phi V\simeq0\), to isolate the recoil instability. The relevant symmetry test is reciprocal driving with equal photon fluxes in the two circulation directions, \(|S_+|^2=|S_-|^2\). To avoid imposing a static optical lattice, the relative pump phase is not fixed on the mechanical locking time, so the conservative standing-wave force averages out. Operationally, this corresponds to mutually incoherent pumps, or to a small relative pump-frequency offset that averages the conservative optical lattice over the mechanical locking time. A fully coherent equal-frequency bidirectional pump contains additional static forces and is not the minimal instability considered here. For a local stability analysis, we consider a uniformly rotating scatterer,
\begin{equation}
\phi(t)=\Omega t,
\qquad
\delta=2m\Omega .
\label{eq:Omega_delta}
\end{equation}
The phase factor \(e^{2im\phi(t)}\) then shifts the frequency of the backscattered field by the rotational Doppler shift \(\delta\)~\cite{lavery2013,speirits2014}. Thus the CW-to-CCW and CCW-to-CW scattering processes are weighted by opposite detunings, \(\Delta-\delta\) and \(\Delta+\delta\). 

\noindent In the weak-scattering limit, the photon number associated with each unperturbed pump is
\begin{equation}
n_0=
\frac{|S|^2}{\gamma^2+\Delta^2},
\qquad
|S_+|=|S_-|\equiv |S|.
\label{eq:n0}
\end{equation}
In this limit, the Doppler-shifted scattering fluxes are \(2\gamma J^2 n_0/[\gamma^2+(\Delta\mp2m\Omega)^2]\), 
up to the sign of the angular momentum transferred. Multiplying the two Doppler-shifted scattering fluxes by the recoil quanta \(\pm2m\hbar\) gives the reciprocal-drive, time-averaged optical torque
\begin{equation}
\tau_{\rm rec}(\Omega)
=
A_m
\left[
\frac{1}{\gamma^2+(\Delta-2m\Omega)^2}
-
\frac{1}{\gamma^2+(\Delta+2m\Omega)^2}
\right].
\label{eq:tau_sym}
\end{equation}
with $A_m = 4m\hbar\gamma J^2 n_0 $. This torque law contains the central feedback mechanism. At \(\Omega=0\), the two reciprocal recoil rates are equal, and the net torque vanishes. At finite \(\Omega\), the rotational Doppler shift makes one scattering channel closer to resonance and the other farther from resonance. The torque is therefore odd in \(\Omega\), and the stability of the reciprocal state is controlled by its slope at the origin.

\paragraph{Chiral instability.---}
The stability of the reciprocal state is set by the linear slope of the reduced torque \(\tau_{\rm rec}(\Omega)\). For small angular velocity,
\begin{equation}
\tau_{\rm rec}(\Omega)
=
\Gamma_{\rm opt}\Omega
+
O(\Omega^3),
\quad
\Gamma_{\rm opt}
=
\frac{
32m^2\hbar\gamma J^2 n_0\Delta
}{
(\gamma^2+\Delta^2)^2
}.
\label{eq:Gamma_opt}
\end{equation}
\begin{figure}[t]
\centering
\includegraphics[width=\columnwidth]{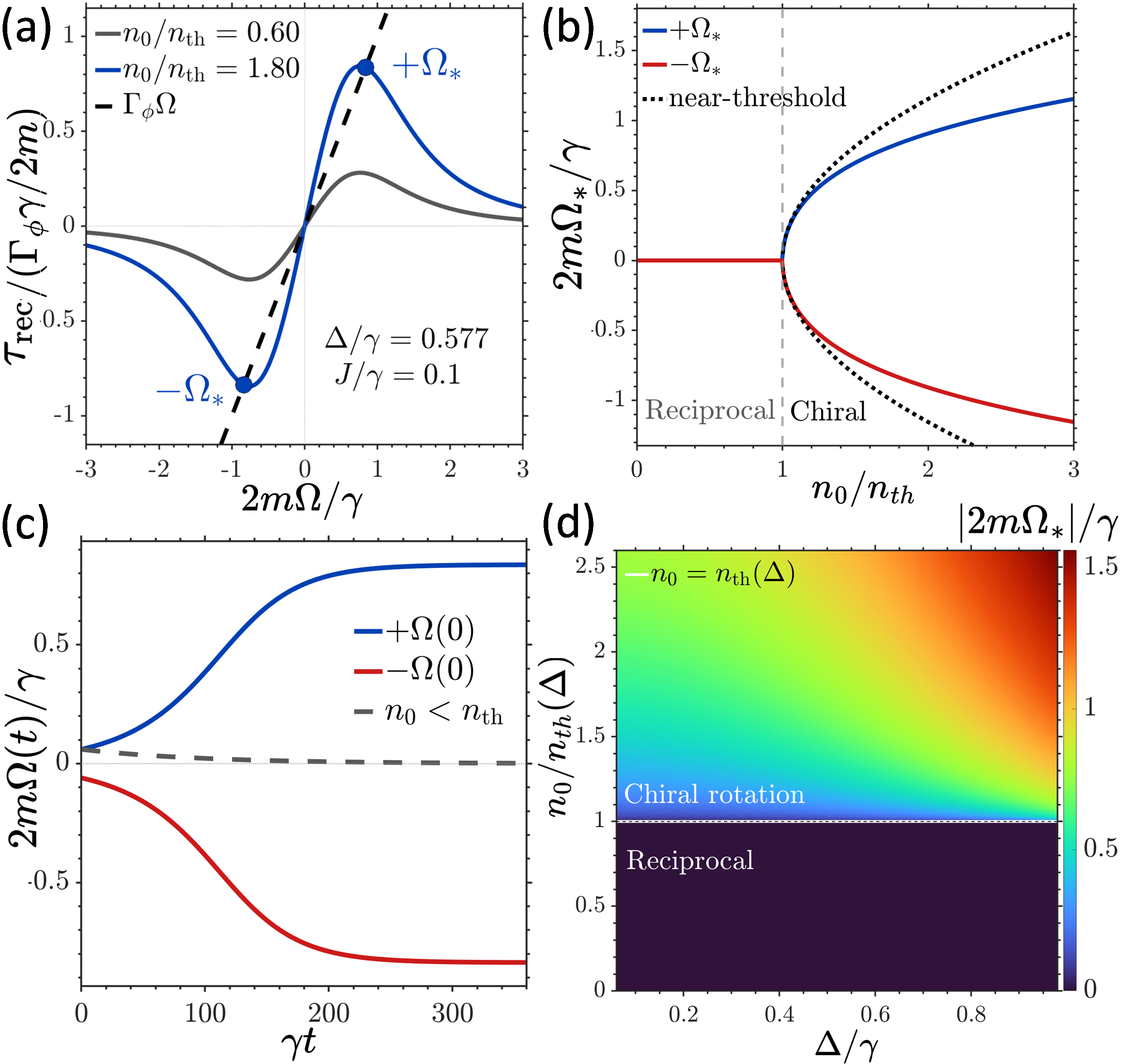}
\caption{
\textbf{Reciprocal angular-recoil instability of a movable WGM backscatterer.} 
(a) Reciprocal-drive recoil torque \(\tau_{\rm rec}(\Omega)\), normalized by
\(\Gamma_\phi\gamma/(2m)\), as a function of the normalized rotational Doppler shift \(2m\Omega/\gamma\), compared with the mechanical damping torque \(\Gamma_\phi\Omega\). For \(n_0<n_{\rm th}\) the only stable solution is \(\Omega=0\), whereas for \(n_0>n_{\rm th}\) two finite intersections \(\Omega=\pm\Omega_*\) appear. Parameters in (a)--(c) are \(\Delta/\gamma=1/\sqrt{3}\simeq0.577\) and \(J/\gamma=0.1\). 
(b) Steady-state angular velocity obtained from \(\Gamma_\phi\Omega=\tau_{\rm rec}(\Omega)\) as a function of the normalized intracavity photon number \(n_0/n_{\rm th}\). The two branches \(\pm\Omega_*\) are related by exchange of the clockwise and counterclockwise directions; the dotted curve shows the mean-field square-root prediction from the local \(Z_2\) normal form.
(c) Time-domain evolution of the reduced rotor equation \(I\dot{\Omega}=\tau_{\rm rec}(\Omega)-\Gamma_\phi\Omega\). Above threshold, infinitesimal-velocity seeds relax to opposite chiral rotations, while below threshold the motion decays back to \(\Omega=0\). 
(d) Instability phase diagram in the \((\Delta,n_0)\) plane, with \(n_0\) normalized by the local threshold \(n_{\rm th}(\Delta)\). The color scale gives the saturated Doppler shift \(|2m\Omega_*|/\gamma\). The white line marks \(n_0=n_{\rm th}(\Delta)\), separating the reciprocal phase from the mechanically chiral rotating phase. 
}
\label{fig:angular_recoil_instability}
\end{figure}
\begin{figure*}[t]
\centering
\includegraphics[width=0.96\textwidth]{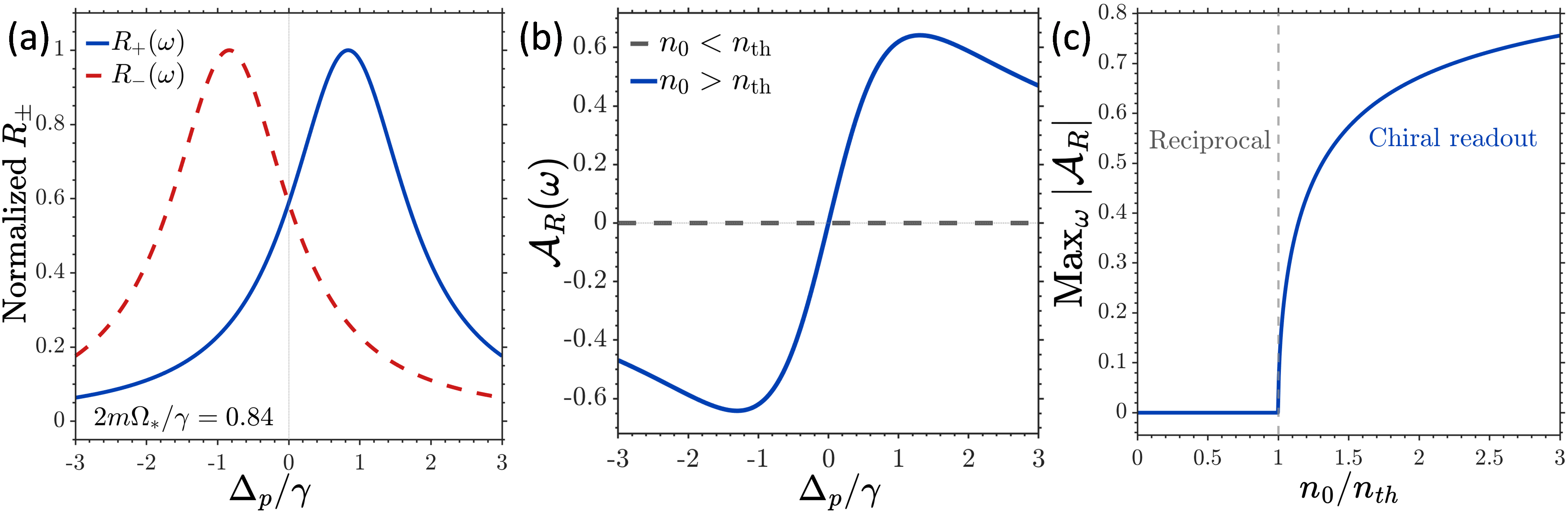}
\caption{
\textbf{Optical readout of the mechanically chiral state.} 
(a) Normalized backscattered spectra \(R_+(\omega)\) and \(R_-(\omega)\), for the two probe directions in the rotating phase.The selected angular velocity \(\Omega_*\) Doppler-shifts the two backscattering channels in opposite directions, producing resonances centered at detunings separated by \(4m\Omega_*\). The plotted case has \(2m\Omega_*/\gamma=0.84\). 
(b) Directional backscattering asymmetry
\(\mathcal A_R(\omega)=[R_+(\omega)-R_-(\omega)]/[R_+(\omega)+R_-(\omega)]\) defined
in Eq.~\eqref{eq:AR}. Below threshold, \(\Omega_*=0\), and the asymmetry vanishes. Above threshold, the Doppler-split spectra produce an odd asymmetry in the probe detuning \(\Delta_p\). 
(c) Maximum asymmetry \(\max_\omega |\mathcal A_R|\) as a function of normalized photon number \(n_0/n_{\rm th}\). The optical readout switches on at the same threshold as the mechanical chiral instability, indicating that the direction-dependent optical response arises from self-selected rotation rather than an externally imposed bias.
}
\label{fig:optical_readout}
\end{figure*}
Thus, the optical recoil contributes a velocity-proportional torque whose sign is controlled by detuning~\cite{kippenberg2005,aspelmeyer2014b}. With the convention \(\Delta=\omega_L-\omega_c\), negative detuning gives ordinary rotational damping, whereas positive detuning gives torque in the direction of the initial motion. Physically, for \(\Omega>0\) and \(\Delta>0\), one recoil channel is shifted closer to resonance while the opposite channel is shifted farther away; the imbalance amplifies the rotation. The linearized angular dynamics are therefore
\begin{equation}
I\dot\Omega
=
-\left(\Gamma_\phi-\Gamma_{\rm opt}\right)\Omega .
\label{eq:linear_stability}
\end{equation}
The nonrotating reciprocal state loses stability when the optical anti-damping exceeds the mechanical damping, $\Gamma_{\rm opt}>\Gamma_\phi$, or, equivalently,
\begin{equation}
n_0>n_{\rm th}
=
\frac{
\Gamma_\phi(\gamma^2+\Delta^2)^2
}{
32m^2\hbar\gamma J^2\Delta
},
\qquad
\Delta>0 .
\label{eq:nth}
\end{equation}
The threshold scales as \(m^{-2}\). This scaling is specific to the angular-recoil mechanism: each backscattered photon transfers \(2m\hbar\), and the rotational Doppler shift that makes the scattering rates velocity dependent is also proportional to \(2m\Omega\). Near threshold, the same expansion gives
\begin{equation}
\tau_{\rm rec}(\Omega)
=
\Gamma_{\rm opt}\Omega
-
u_{\rm opt}\Omega^3
+
O(\Omega^5),
\label{eq:cubic_torque}
\end{equation}
with $u_{\rm opt}=\left(8\Gamma_{\rm opt}m^2(\gamma^2-\Delta^2)\right)/(\gamma^2+\Delta^2)^2$. 

\noindent The rotor dynamics, therefore, reduces to
\begin{equation}
I\dot{\Omega}
=
r\Omega-u_{\rm opt}\Omega^3+O(\Omega^5),
\qquad
r=\Gamma_{\rm opt}-\Gamma_\phi .
\label{eq:normal_form}
\end{equation}
For \(0<\Delta<\gamma\), one has \(u_{\rm opt}>0\), so the instability is supercritical. Thus, the transition is a mean-field \(Z_2\) bifurcation of the velocity order parameter~\cite{guckenheimer1983,cross1993}: the relaxation rate below threshold,
\((\Gamma_\phi-\Gamma_{\rm opt})/I\), vanishes at onset, while above threshold the stable states satisfy \(|\Omega_*|\propto r^{1/2}\). The cubic term in Eq.~\eqref{eq:normal_form} is the local expression of a simple physical saturation. The Doppler shift that creates the recoil imbalance also detunes the resonant scattering channel at larger \(|\Omega|\). The finite-velocity states are determined by $\Gamma_\phi\Omega=\tau_{\rm rec}(\Omega)$. Close to threshold, writing \(\mu=n_0/n_{\rm th}\), the supercritical branch in the regime \(0<\Delta<\gamma\) is
\begin{equation}
\Omega_*
\simeq
\frac{
\gamma^2+\Delta^2
}{
2m\sqrt{2(\gamma^2-\Delta^2)}
}
\sqrt{\mu-1}.
\label{eq:Omega_star}
\end{equation}
The two solutions \(\Omega=\pm\Omega_*\) are exchanged by reversing the two circulation directions. 

\noindent Reciprocal pumping, therefore, leaves the two handednesses degenerate but makes the zero-velocity state unstable above threshold. Within the same weak-scattering theory, the threshold is minimized at $\Delta_{\rm opt}=\gamma/\sqrt{3}$. This gives the local pitchfork structure of the chiral instability \cite{strogatz2018a}.
\paragraph{Optical signature and scope.---}
The chiral mechanical state has a direct optical readout. Once the system selects a rotation \(\Omega_*\), backscattering from a weak probe is no longer resonant at the same detuning in the two directions. A probe incident in the \(+\) direction couples to the opposite circulation through a sideband shifted by \(-2m\Omega_*\), whereas a probe incident in the \(-\) direction couples through the sideband shifted by \(+2m\Omega_*\)~\cite{lavery2013,speirits2014, hatifi2026c}. The backscattered field provides the most direct probe of the rotational Doppler splitting. For a weak probe launched into the \(+\) or \(-\) circulation channel, the amplitude scattered into the opposite channel is, in the same coupled-mode approximation,
\begin{equation}
r_\pm(\omega)
=
\frac{
-i\kappa_{\rm ex}J
}{
\left(\gamma-i\Delta_p\right)
\left[\gamma-i(\Delta_p\mp2m\Omega_*)\right]
+
J^2
},
\label{eq:rpm}
\end{equation}
where \(\Delta_p=\omega-\omega_c\) is the probe detuning and \(\kappa_{\rm ex}\) is the external coupling rate. The corresponding backscattered spectra are simply given by $R_\pm(\omega)=|r_\pm(\omega)|^2$. In the weak-scattering limit \(J\ll\gamma\), Eq.~\eqref{eq:rpm} gives
\begin{equation}
R_\pm(\omega)
\simeq
\frac{
\kappa_{\rm ex}^2 J^2
}{
\left(\gamma^2+\Delta_p^2\right)
\left[
\gamma^2+
\left(\Delta_p\mp2m\Omega_*\right)^2
\right]
}.
\label{eq:Rpm_weak}
\end{equation}
Thus \(R_+(\omega)\) and \(R_-(\omega)\) coincide below threshold, where \(\Omega_*=0\), and split once the system selects a finite rotation. We quantify this directional readout by the normalized backscattering asymmetry
\begin{equation}
\mathcal A_R(\omega)
=
\frac{R_+(\omega)-R_-(\omega)}
{R_+(\omega)+R_-(\omega)} .
\label{eq:AR}
\end{equation}
This quantity vanishes in the reciprocal phase and becomes finite only in the mechanically chiral phase ~\cite{hatifi2026b,hatifi2026c}. The same sideband elimination gives the same-frequency through response
\begin{equation}
t_\pm(\omega)
=
1-
\frac{\kappa_{\rm ex}}
{
\gamma-i\Delta_p
+
\dfrac{J^2}{\gamma-i(\Delta_p\mp 2m\Omega_*)}
}.
\label{eq:tpm}
\end{equation}
Consequently, $T_+(\omega)=|t_+(\omega)|^2 \neq T_-(\omega)=|t_-(\omega)|^2$
whenever \(\Omega_*\neq0\). The direction dependence is therefore a readout of the mechanically selected handedness. It is not imposed by an external rotation; it appears only after the recoil instability has generated the chiral state. This direction-dependent weak-probe response should not be confused with ideal isolation in the sense of nonlinear optical isolators~\cite{shi2015}. This effect should be separated from nearby mechanisms. In static-scatterer WGM splitting, \(\phi\) is a fixed parameter, and the recoil channel cannot destabilize a mechanical velocity~\cite{gorodetsky2000,mazzei2007}. In one-sided optical transport, the drive already provides a preferred direction of flow~\cite{arnold2009}. In externally spinning WGM resonators, \(\Omega\) is an imposed control parameter~\cite{maayani2018,jing2018}. Kerr ring resonators provide the closest symmetry-breaking analogy, because bidirectional driving can select one of two optical circulation states~\cite{delbino2017}. Here, the order parameter is different; the selected quantity is the mechanical angular velocity, and the optical asymmetry follows from its self-generated rotational Doppler shift.

\paragraph{Scope of the minimal theory.---}
The minimal theory is designed to isolate angular recoil as the source of chirality. We remove effects that would select a direction independently of the recoil dynamics, including static standing-wave trapping, one-sided optical propulsion, externally imposed rotation, and strong-scattering multimode dynamics. 
\\ 

\noindent The analytically controlled regime is a single WGM doublet with weak coherent backscattering, Markovian optical loss, a classical angular coordinate, negligible static pinning, and reciprocal pumps whose relative phases are averaged so that no stationary optical lattice is imposed. Under these conditions, chirality can arise only from the velocity-dependent imbalance between the two recoil channels. They remove ordinary standing-wave trapping and one-sided optical propulsion, leaving only the velocity-dependent imbalance between the two recoil channels. The result is therefore a mechanism-level statement rather than an optimized device proposal. Viewed more broadly, the instability belongs to the class of driven nonlinear hybrid systems in which gain or parametric backaction converts pump energy into autonomous mechanical motion or structured emission~\cite{kippenberg2005,aspelmeyer2014b,kani2025b,hatifi2026a}; here, however, the gain channel is angular recoil from circulation-changing WGM backscattering rather than a Kerr, magnonic, or conventional radiation-pressure nonlinearity. A concrete implementation must still minimize the moment of inertia and angular damping of the movable scatterer while preserving a high optical quality factor, favoring evanescently coupled nanoscale perturbations or suspended angular elements over a macroscopic membrane placed in the optical path~\cite{vahala2003,arnold2009,foreman2015}. Finite pinning, pump imbalance, coherent-lattice forces, noise-induced switching between \(\pm\Omega_*\)~\cite{hanggi1990}, and corrections beyond \(J\ll\gamma\) are natural extensions for platform-specific modeling.

\paragraph{Conclusion.---}
We have shown that WGM mode splitting can become an active angular-momentum backaction mechanism when the scatterer angle is promoted to a mechanical coordinate. In this setting, each circulation-changing photon transfers a recoil \(\pm2m\hbar\), while rotation Doppler-shifts the two reciprocal scattering rates in opposite directions. The resulting feedback converts reciprocal optical driving into negative angular friction for the appropriate detuning, destabilizing zero rotation and selecting one of two mechanically chiral states. The onset is a dynamical critical point of the reciprocal driven state, rather than an externally biased motor transition. Its optical signature is a direction-dependent weak-probe response generated by the self-selected mechanical velocity, not by externally spinning the resonator. Thus, a single dynamical WGM backscatterer turns passive defect-induced mode splitting into a minimal mechanism for autonomous chiral optomechanics.

\begin{acknowledgments}
\paragraph{Acknowledgements—}
The author would like to thank Brian Stout, Jason Twamley, Thomas Durt, and Branko Kolaric for stimulating discussions. 
\end{acknowledgments}

\bibliographystyle{apsrev4-2}
%

\end{document}